\documentclass[10pt]{amsart}
\usepackage{graphicx}
\usepackage{amsmath}
\usepackage{amsthm}
\usepackage{amsfonts}
\usepackage{amssymb}
\usepackage{cite}
\usepackage{enumerate}

\usepackage{epic}
\usepackage{pstricks}

\theoremstyle{plain}
\newtheorem{theorem}{Theorem}[section]
\newtheorem{proposition}[theorem]{Proposition}

\newtheorem{lemma}[theorem]{Lemma}

\newtheorem{definition}[theorem]{Definition}
\newtheorem{remark}[theorem]{Remark}
\newtheorem*{remark*}{Remark}

\newcommand{\uu}{\vec{u}}
\newcommand{\vv}{\vec{v}}
\renewcommand{\v}[2]{\overrightarrow{V_{#1}V_{#2}}}

\newcommand{\pp}{\uparrow\!\uparrow}

\newcommand{\intr}[2]{\overline{#1,#2}}

\newcommand{\De}{\Delta}
\newcommand{\al}{\alpha}
\newcommand{\be}{\beta}
\newcommand{\la}{\lambda}

\renewcommand{\th}{\theta}

\newcommand{\ga}{\gamma}

\newcommand{\vpi}{\varphi}

\renewcommand{\le}{\leqslant}
\renewcommand{\ge}{\geqslant}

\newcommand{\R}{\mathbb{R}}
\newcommand{\Z}{\mathbb{Z}}

\renewcommand{\P}{\mathcal{P}}
\newcommand{\Q}{\mathcal{Q}}

\newcommand{\conv}{\operatorname{conv}}
\newcommand{\edg}{\operatorname{edg}}
\newcommand{\ri}{\operatorname{ri}\,}
\newcommand{\sign}{\operatorname{sign}}

\begin{document}


\title[An $O(n)$ Polygon Convexity Test]
{Polygon Convexity: 
Another $O(n)$ Test} 


\author{Iosif Pinelis}  
                   

\date{\today; file convex-poly/test/mehlhorn/\jobname.tex}

\begin{abstract}
An $n$-gon is defined as a sequence $\P=(V_0,\dots,V_{n-1})$ of $n$ points on the plane. 
An $n$-gon $\P$ is said to be convex if the boundary of the convex hull of the set $\{V_0,\dots,V_{n-1}\}$ of the vertices of $\P$ coincides with the union of the edges $[V_0,V_1],\dots,[V_{n-1},V_0]$; if at that no three vertices of $\P$ are collinear then $\P$ is called strictly convex. We prove that an $n$-gon $\P$ with $n\ge3$ is strictly convex if and only if a cyclic shift of the sequence $(\al_0,\dots,\al_{n-1})\in[0,2\pi)^n$ of the angles between the $x$-axis and the vectors $\v01,\dots,\v{n-1}0$ is strictly monotone.

A ``non-strict'' version of this result is also proved.
\end{abstract}



\maketitle

\section{Definitions and results}\label{results}

A {\em polygon} is defined in this paper as any finite sequence of points (or, interchangeably, vectors) on the Euclidean plane $\R^2$; the same definition was used in \cite{jgeom,elimin,test,szek}. Let here $\P:=(V_0,\dots,V_{n-1})$ be a polygon, which is sequence of $n$ points; such a polygon is also called an $n$-gon. 
The points $V_0,\dots,V_{n-1}$ are called the {\em vertices} of $\P$.
The smallest value that one may allow for the integer $n$ is $0$, corresponding to a polygon with no vertices, that is, to the sequence $()$ of length $0$. 
The segments, or closed intervals,
$$[V_i,V_{i+1}]:=\conv\{V_i,V_{i+1}\}\quad\text{for}\ i\in\{0,\dots,n-1\}$$
are called the {\em edges} of polygon $\P$, where 
$$V_n:=V_0.$$
The symbol $\conv$ denotes, as usual, the convex hull \cite[page 12]{rock}.
Note that, if $V_i=V_{i+1}$, then the edge $[V_i,V_{i+1}]$ is a singleton set. 
For any two points $U$ and $V$ in $\R^2$, let $[U,V]:=\conv\{U,V\}$, $[U,V):=[U,V]\setminus\{V\}$, and $(U,V):=[U,V]\setminus\{U,V\}$, so that $(U,V)=\ri[U,V]$, the relative interior of $[U,V]$.
Let us define the convex hull and dimension of polygon $\P$ as, respectively, the convex hull and dimension of the set of its vertices: $\conv\P:=\conv\{V_0,\dots,V_{n-1}\}$ and $\dim\P:=\dim\{V_0,\dots,V_{n-1}\}=\dim\conv\P$.
In general, our terminology corresponds to that in \cite{rock}. 

In the sequel, we also use the notation
$$\intr km:=\{i\in\Z\colon k\le i\le m\},$$
where $\Z$ is the set of all integers; in particular, $\intr km$ is empty if $m<k$. 

Given the above notion of the polygon, a {\em convex polygon} can be defined as a polygon $\P$ such that the union of the edges of $\P$ coincides with the boundary $\partial\conv\P$ of the convex hull $\conv\P$ of $\P$; cf. e.g. \cite[page 5]{yaglom}. 
Thus, one has

\begin{definition}\label{def:conv}
A polygon $\P=(V_0,\dots,V_{n-1})$ is {\em convex} if 
$$\bigcup_{i\in\intr0{n-1}}[V_i,V_{i+1}]=\partial\conv\P.$$
\end{definition}

Let us emphasize that a polygon in this paper is a sequence and therefore ordered. In particular, even if all the vertices $V_0,\dots,V_{n-1}$ of a polygon $\P=(V_0,\dots,V_{n-1})$ are the extreme points of the convex hull of $\P$, it does not necessarily follow that $\P$ is convex. For example, consider the points $V_0=(0,0)$, $V_1=(1,0)$, $V_2=(1,1)$, and $V_3=(0,1)$. Then polygon $(V_0,V_1,V_2,V_3)$ is convex, while polygon $(V_0,V_2,V_1,V_3)$ is not.


\begin{definition}\label{def:strict}
Let us say that a polygon $\P=(V_0,\dots,V_{n-1})$ is 
\begin{itemize}
\item
{\em locally-ordinary} -- if 
for any $i$ in the set $\intr0{n-1}$ the vertices $V_i$ and $V_{i+1}$ are distinct;
\item
{\em ordinary} -- if 
for any two distinct $i$ and $j$ in $\intr0{n-1}$ the vertices $V_i$ and $V_j$ are distinct;
\item
{\em locally-strict} -- if 
for any $i$ in $\intr0{n-1}$ the vertices $V_{i-1}$, $V_i$, and $V_{i+1}$ are non-collinear, where $V_{-1}:=V_{n-1}$;
\item
{\em quasi-strict} -- if
any two adjacent vertices of $\P$ are not collinear with any other vertex of $\P$ or, more formally, 
if for any $i\in\intr0{n-1}$ and any $j\in\intr0{n-1}\setminus\{i,i\oplus1\}$ the points $V_i$, $V_{i\oplus1}$, and $V_j$ are non-collinear, where 
$$i\oplus1:=
\begin{cases}
i+1 &\text{if}\ i\in\intr0{n-2},\\
0 &\text{if}\ i=n-1; 
\end{cases}
$$ 
\item
{\em strict} -- if 
for any three distinct $i$, $j$, and $k$ in $\intr0{n-1}$ the vertices $V_i$, $V_j$, and $V_k$ are non-collinear;
\item
{\em locally-simple} -- if 
for any $i$ in $\intr0{n-1}$ one has $[V_i,V_{i+1})\cap[V_{i+1},V_{i+2})=\emptyset$, where $V_{n+1}:=V_1$;
\item
{\em simple} -- if 
for any two distinct $i$ and $j$ in $\intr0{n-1}$ one has $[V_i,V_{i+1})\cap[V_j,V_{j+1})=\emptyset$;
\item
{\em locally-ordinarily convex} -- if 
$\P$ is locally-ordinary and convex;
similarly can be defined \emph{ordinarily convex}, \dots, \emph{simply convex} polygons.
\end{itemize}
\end{definition}


\begin{remark}\label{rem:simpl->ord}
Any (locally-)simple polygon is (locally-)ordinary, since $\{V_i\}\cap\{V_j\}\subseteq[V_i,V_{i+1})\cap[V_j,V_{j+1})$. 
\end{remark}

We shall make use of the following result given in \cite{elimin}. 
If $\P=(V_0,\dots,V_{n-1})$ is a polygon, let us refer to any subsequence $(V_{i_0},\dots,V_{i_{m-1}})$ of $\P$, with $0\le i_0<\dots<i_{m-1}\le n-1$,  as a {\em sub-polygon} or, more specifically, 
as a {\em sub-$m$-gon} of $\P$.

\begin{theorem}\label{cor:sub-polygon}
\emph{\cite[Corollary~1.17]{elimin}}
If $\P=(V_0,\dots,V_{n-1})$ is an ordinarily convex polygon, then any sub-polygon of $\P$ is so; in particular, then the sub-$(n-1)$-gons $\P^{(i)}:=(V_0,\dots,V_{i-1},V_{i+1},\dots,V_{n-1})$ of $\P$ are ordinarily convex, for all $i\in\intr0{n-1}$. 
\end{theorem}

For a polygon $\P=(V_0,\dots,V_{n-1})$, let $x_i$ and $y_i$ denote the coordinates of its vertices $V_i$, so that
\begin{equation}\label{eq:xiyi}
	V_i=(x_i,y_i)\quad\text{for}\ i\in\intr0{n-1}.
\end{equation}
Introduce the determinants
\begin{equation}\label{eq:De}
\De_{\al,i,j}:=
\left|
\begin{matrix}
1&x_\al&y_\al \\
1&x_i&y_i \\
1&x_j&y_j 
\end{matrix}\right|
\end{equation}
for $\al$, $i$, and $j$ in the set $\intr0{n-1}$. 
Let then
$$
\begin{aligned}
a_i&:=\sign\De_{i+1,i-1,i}=\sign\De_{i-1,i,i+1};\\
b_i&:=\sign\De_{0,i-1,i};\\
c_i&:=\sign\De_{i,0,1}=\sign\De_{0,1,i}.
\end{aligned}
$$

The following theorem is the main result of \cite{test}, which provides an $O(n)$ test of the strict convexity of a polygon.

\begin{theorem}\label{th:calculation}
\emph{\cite{test}}\ An $n$-gon $\P=(V_0,\dots,V_{n-1})$ with $n\ge4$ is strictly convex if and only if conditions
\begin{equation}\label{eq:conds}
	\begin{aligned}
a_i b_i& >0,\\
a_i b_{i+1}& >0,\\
c_i c_{i+1}& >0
\end{aligned}
\end{equation}
hold for all
$$i\in\intr2{n-2}.$$
\end{theorem}

\begin{proposition}\label{prop:minimality}
\emph{\cite{test}}\ None of the $3(n-3)$ conditions in Theorem~\ref{th:calculation} can be omitted without (the ``if" part of) Theorem~\ref{th:calculation} ceasing to hold. 
\end{proposition}

Thus, the test given by Theorem~\ref{th:calculation} is exactly minimal.

\begin{remark}\label{rem:minimality}
\emph{\cite{test}}\ Adding to the $3(n-3)$ conditions \eqref{eq:conds} in Theorem~\ref{th:calculation} the equality $b_2=c_2$, which trivially holds for any polygon (convex or not), one can rewrite \eqref{eq:conds} as the following system of $3(n-3)+1$ equations and one inequality:
$$
	\begin{aligned}
&a_2=\dots=a_{n-2}\\
=&b_2=\dots=b_{n-2}=b_{n-1}\\
=&c_2=\dots=c_{n-2}=c_{n-1}\ne0.
\end{aligned}
$$
\end{remark}

These results were used in \cite{szek}.

For any vector $\vv=(x,y)\in\R^2$ with $r:=|\vv|:=\sqrt{x^2+y^2}\ne0$, define the (angle) argument of $\vv$ as usual, by the formula
\begin{equation}\label{eq:arg-def}
\arg\vv=\psi\Longleftrightarrow
\big(0\le\psi<2\pi\ \&\ x=r\cos\psi\ \&\ y=r\sin\psi \big),	
\end{equation}
so that, for each nonzero vector $\vv\in\R^2$, the ``angle'' $\arg\vv$ is a uniquely defined number in the interval $[0,2\pi)$.
Moreover, 
\begin{gather}\label{eq:arg}
\arg\vv=
\begin{cases} 
\arccos\tfrac xr &\ \text{ if } \vv\in H_-,\\
2\pi-\arccos\tfrac xr &\ \text{ if } \vv\in H_+,\\
\end{cases} 
\intertext{where $\arccos$ is the branch of the inverse function $\cos^{-1}$ with values in the interval $[0,\pi]$ and} 
\begin{aligned}
H_-&:=\{(x,y)\in\R^2\colon y>0\ \text{or}\ (y=0\ \&\ x>0)\}, \\ 
H_+&:=\{(x,y)\in\R^2\colon y<0\ \text{or}\ (y=0\ \&\ x<0)\}; \notag
\end{aligned}
\end{gather}
note that $H_-\cup H_+=\R^2\setminus\{\vec0\}$ and $H_-\cap H_+=\emptyset$.

For any locally-ordinary polygon $\P=(V_0,\dots,V_{n-1})$, introduce also the sequence of the angle arguments the edge-vectors $\v01,\dots,\v{n-1}n$ of $\P$, by the formula 
$$\arg\P:=(\arg\v01,\dots,\arg\v{n-1}n).$$

For any two nonzero vectors $\uu$ and $\vv$ in $\R^2$, let us write $$\uu<\vv\quad\text{iff}\quad\arg\uu<\arg\vv;$$ 
similarly defined is the relation 
$>$ 
on $\R^2\setminus\{\vec0\}$. 

\begin{remark}\label{rem:angles}
Let $\uu=(s,t)$ and $\vv=(x,y)$ be any two vectors in $\R^2\setminus\{\vec0\}$.
Then, using \eqref{eq:arg}, it is elementary but somewhat tedious to check that 
\begin{gather}\label{eq:mehl-order}
	\uu<\vv\iff
	\begin{cases}
\uu\in H_-\ \& \ \vv\in H_+ &\ \text{\em or}\\
\uu\in H_-\ \& \ \vv\in H_-\ \&\ 
\De>0 &\ \text{\em or}\\
\uu\in H_+\ \& \ \vv\in H_+\ \&\ 
\De>0,
	\end{cases} 
\intertext{where}
\De:=\left|
\begin{matrix}
1&0&0 \\
1&s&t \\
1&x&y 
\end{matrix}\right|	
=\left|
\begin{matrix}
s&t \\
x&y 
\end{matrix}\right|	
=sy-tx. \notag
\end{gather}
Under the additional condition that $\uu$ and $\vv$ are non-collinear, it follows that either $\uu<\vv$ or $\vv<\uu$:
$$\De\ne0\implies(\uu<\vv\ \text{or}\ \vv<\uu).$$
Note also that
\begin{equation}\label{eq:mehl-order-simpl}
	\uu<\vv\implies
	(y\le0\le t\ \text{\emph{or}}\ \De>0).
\end{equation}
\end{remark}

\begin{definition}\label{def:incr}
Let us say that a locally-ordinary polygon $\P=(V_0,\dots,V_{n-1})$ with $\arg\P=:(\al_0,\dots,\al_{n-1})$ is 
\begin{itemize}
\item
{\em increasing} -- if 
the sequence $\arg\P$ is increasing: $\al_0<\dots<\al_{n-1}$;
\item
{\em decreasing} -- if $\al_0>\dots>\al_{n-1}$;
\item
\emph{cyclically increasing}  or, briefly, \emph{c-increasing} -- if $$\al_k<\dots<\al_{n-1}<\al_0<\dots<\al_{k-1},$$
for some $k\in\intr0{n-1}$;
(if $k=0$ then this chain of inequalities is supposed to read simply as $\al_0<\dots<\al_{n-1}$, in which case polygon $\P$ will be increasing);
\item
\emph{cyclically decreasing}  or, briefly, \emph{c-decreasing} -- similarly, if $$\al_k>\dots>\al_{n-1}>\al_0>\dots>\al_{k-1},$$
for some $k\in\intr0{n-1}$;
\item
\emph{cyclically strictly monotone}  or, briefly, \emph{c-strictly monotone} -- if $\P$ is either c-increasing or c-decreasing.
\end{itemize}
The notions of nondecreasing, nonincreasing, c-nondecreasing, c-nonincreasing, and c-monotone polygons are defined similarly, with signs $\le$ and $\ge$ replacing $<$ and $>$, respectively.
\end{definition}

For any $k\in\intr0{n-1}$, define the cyclic shift (or, briefly, c-shift) $\th^k$ of the sequence $(u_0,\dots,u_{n-1})$ of any objects $u_0,\dots,u_{n-1}$ by the formula
$$(u_0,\dots,u_{n-1})\th^k:=(u_k,\dots,u_{n-1},u_0,\dots,u_{k-1}).$$

\begin{remark}\label{rem:c-shift}
For any polygon $\P=(V_0,\dots,V_{n-1})$ and any $k\in\intr0{n-1}$,  one has $\arg(\P\th^k)=(\arg\P)\th^k$. It follows that $\P$ is c-increasing iff $\P$ is a c-shift of an increasing polygon iff a c-shift of $\P$ is increasing; similarly for ``decreasing'' vs.\  ``c-decreasing'' and for other such pairs of terms. Also, all c-shifts preserve the polygon convexity and all properties defined in Definition~\ref{def:strict} as well as all the ``cyclic'' properties defined in Definition~\ref{def:incr}: being c-increasing, being c-increasing,\dots. 
\end{remark}

For any transformation $T\colon\R^2\to\R^2$ and any polygon $\P=(V_0,\dots,V_{n-1})$, define the corresponding transformation of $\P$ as the polygon $T\P:=(TV_0,\dots,TV_{n-1})$. A rotation is any orthogonal (and hence linear) transformation with determinant $1$; any rotation can be represented as the linear transformation $R_\al$ with matrix $[\begin{smallmatrix}\cos\al&-\sin\al\\ \sin\al&\cos\al\end{smallmatrix}]$ for some real number $\al$, so that $R_\al[\begin{smallmatrix}1\\0\end{smallmatrix}]=[\begin{smallmatrix}\cos\al\\ \sin\al\end{smallmatrix}]$.
The reflection is denoted here by $R$ and defined by the formula $\R^2\ni(x,y)\mapsto R(x,y):=(x,-y)$. Any orthogonal transformation can be represented as $R_\al R$ (as well as $RR_\be$) for appropriate $\al$ and $\be$.
A homothetical transformation is understood here as one of the form $\R^2\ni\vv\to\la\vv$ for some $\la>0$. 

\begin{proposition}\label{prop:preserv}
The properties of being convex; 
locally-ordinary; ordinary; locally-strict; quasi-strict; 
strict; locally-simple; and simple 
are each preserved for any polygon under any
nonsingular affine transformation. 
The properties of being c-increasing, c-decreasing, c-nondecreasing, and c-nonincreasing
are each preserved under any
rotation or any homothetical transformation or any parallel translation. 
The properties of being c-increasing and c-decreasing are interchanged under the reflection.
\end{proposition}

All the necessary proofs are deferred to Section~\ref{proofs}.

The following theorem is one of the main results of this paper.

\begin{theorem}\label{th:angl-test-strict}
An $n$-gon with $n\ge3$ is strictly convex iff it is c-strictly monotone. 
\end{theorem}

\begin{remark}\label{rem:a-text-strict}
Any $n$-gon with $n\le1$ is, trivially, both strictly convex and c-strictly monotone. Any $2$-gon is, trivially, strictly convex; however, a $2$-gon is c-strictly monotone only if it is locally-ordinary (and hence ordinary). It is easy to see that any strict $3$-gon is c-strictly monotone, so that Theorem~\ref{th:angl-test-strict} is trivial for $n=3$. 
Note also that an $n$-gon is both c-increasing and c-decreasing iff it is locally-ordinary and $n=2$.
\end{remark}

Theorem~\ref{th:angl-test-strict} is complemented by the following proposition, which will also be of use in the proof of 
Theorem~\ref{th:angl-test-strict}.

\begin{proposition}\label{prop:strict}
For any 
$n$-gon 
$\P$ 
with $n\ge3$ 
the following statements are equivalent to one another:

\renewcommand\theenumi {\emph{(\Roman{enumi})} }
\renewcommand\labelenumi {\theenumi}

\begin{enumerate}
\item
$\P$ is ordinary and locally-strictly convex;
\item
$\P$ is quasi-strictly convex;
\item
$\P$ is strictly convex.
\end{enumerate}
\end{proposition}

\begin{remark*}
The conditions in Proposition~\ref{prop:strict} that $n\ge3$ and $\P$ is ordinary cannot be dropped. Indeed, all $2$-gons are strictly convex, but not all of them are ordinary. On the other hand, the polygon $(V_0,V_1,V_2,V_0,V_1,V_2)$ is locally-strictly convex if the points $V_0,V_1,V_2$ are non-collinear, but it is not ordinary and not strictly convex.
\end{remark*}

The following theorem is a ``non-strict"' counterpart of Theorem~\ref{th:angl-test-strict}.

\renewcommand\theenumi {\emph{(\Roman{enumi})} }
\renewcommand\labelenumi {\theenumi}

\begin{theorem}\label{th:angl-test-nonstrict}
For any
$n$-gon $\P$ with $n\ge3$ the following statements are equivalent to one another
\begin{enumerate}
\item
$\P$ is ordinary and locally-simply convex;
\item
$\P$ is simply convex;
\item
$\P$ is c-monotone and $\dim\P=2$.
\end{enumerate}
\end{theorem}

\begin{remark*}
None of the following conditions: (i) $n\ge3$, (ii) $\P$ is ordinary, or (iii) $\dim\P=2$  in Theorem~\ref{th:angl-test-nonstrict} can be dropped. Indeed, (i) all $2$-gons are simply convex but not all of them are ordinary; (ii) for any three non-collinear points $V_0,V_1,V_2$, the $6$-gon $(V_0,V_1,V_2,V_0,V_1,V_2)$ is of dimension $2$ and locally-simply convex but not simply convex or c-monotone (or ordinary); (iii) for any three distinct collinear points $V_0,V_1,V_2$, the $3$-gon $(V_0,V_1,V_2)$ is c-monotone and ordinary but not simply convex or locally-simply convex (or of dimension $2$).
\end{remark*}

A suggestion to use c-strict monotonicity to test for polygon convexity was given in \cite{mehl}, without a proof. A result, similar to Theorem~\ref{th:angl-test-nonstrict} was presented in \cite[Lemma 5 in Section 10.3]{leda}, with a very brief, heuristic proof.

\section{Proofs}\label{proofs}

\begin{proof}[Proof of Proposition \ref{prop:preserv}]
Suppose that an $n$-gon $\P$ with $(\al_0,\dots,\al_{n-1}):=\arg(\P)$ is c-increasing, that is, 
\begin{equation}
	\al_k<\dots<\al_{n-1}<\al_0<\dots<\al_{k-1},\tag{Incr$_k$}
\end{equation}
for some $k\in\intr0{n-1}$. Let $(\be_0,\dots,\be_{n-1}):=\arg(R\P)$, the argument sequence of the reflected polygon $R\P$. Then $\be_i=2\pi-\al_i$ for all $i\ne k$, while $\be_k=2\pi-\al_k$ if $\al_k\ne0$ and $\be_k=0$ if $\al_k=0$. 
It follows that
\begin{equation}
	\be_k>\dots>\be_{n-1}>\be_0>\dots>\be_{k-1}\tag{Decr$_k$}
\end{equation}
if $\al_k\ne0$, and the $\be_i$'s satisfy (Decr$_{k\oplus1}$) if $\al_k=0$, where $k\oplus1:=k+1$ if $k\in\intr0{n-2}$ and $k\oplus1:=0$ if $k=n-1$. 

Similarly, if $\arg\P=:(\be_0,\dots,\be_{n-1})$ satisfies condition (Decr$_k$) for some $k\in\intr0{n-1}$, then $(\al_0,\dots,\al_{n-1}):=\arg(R\P)$ satisfies condition (Incr$_k$) if $\al_{k-1}\ne0$, and the $\al_i$'s satisfy (Incr$_{k\ominus1}$) if $\al_{k-1}=0$, where $k\ominus1:=k-1$ if $k\in\intr1{n-1}$ and $k\oplus1:=n-1$ if $k=0$. 

Thus, reflection $R$ interchanges the properties of being c-increasing and c-decreasing.

Let us now verify the preservation of the c-increasing property under any rotation $R_\al$. W.l.o.g., $0\le\al<2\pi$. 
Suppose again that an $n$-gon $\P$ with $(\al_0,\dots,\al_{n-1}):=\arg(\P)$ satisfies condition (Incr$_k$). 
Then the c-shift $\Q:=\P\th^k:=(V_k,\dots,V_{n-1},\break
V_0,\dots,V_{k-1})$ of polygon $\P$ with 
$(\be_0,\dots,\be_{n-1}):=\arg(\Q)=
(\al_k,\dots,\al_{n-1},\break
\al_0,\dots,\al_{k-1})$ is an increasing $n$-gon. 
Let $(\psi_0,\dots,\psi_{n-1}):=\arg(R_\al\Q)$. 
Let $J:=\{i\in\intr0{n-1}\colon\be_i+\al\ge2\pi\}$, and let $j:=\min J$ if $J\ne\emptyset$ and $j:=n$ if $J=\emptyset$.
Then 
$\psi_i:=\be_i+\al$ for $i\in\intr0{j-1}$ and $\psi_i:=\be_i+\al-2\pi$ for $i\in\intr j{n-1}$. 
Hence, the sequence $\arg(R_\al\Q\th^j)=:(\vpi_0,\dots,\vpi_{n-1})$ is increasing, where $\vpi_i:=\be_{i+j}+\al-2\pi$ for $i\in\intr0{n-j-1}$, and $\vpi_i:=\be_{i+j-n}+\al$ for $i\in\intr{n-j}{n-1}$. 
Thus, the cyclic permutation $R_\al\P\th^m=R_\al\P\th^{k+j}=R_\al\Q\th^j$ of polygon $R_\al\P$ is increasing, where $m:=k+j$ if $k+j<n$ and $m:=k+j-n$ if $k+j\ge n$.
Thus, $R_\al\P$ is c-increasing.

The preservation of the c-decreasing 
property under any rotation is verified quite similarly. 

The other claims stated in Proposition \ref{prop:preserv} are only easier to check.
\end{proof}

\begin{proof}[Proof of Theorem \ref{th:angl-test-strict}]
Let $\P=(V_0,\dots,V_{n-1})$ be an $n$-gon with $n\ge3$, vertices $V_i=:(x_i,y_i)$, and argument $(\al_0,\dots,\al_{n-1}):=\arg\P$.
In view of Proposition~\ref{prop:preserv}, the rotation $R_{2\pi-\al_0}$ and any homothetical transformation will preserve both the convexity and c-monotonicity properties of $\P$. Therefore,
assume without loss of generality (w.l.o.g.) that 
$\al_0=0
$ and, moreover, $V_0=(0,0)$ and $V_1=(1,0)$.

\textbf{``If''}\quad  
When proving this part, assume w.l.o.g.\ that $\P$ is c-increasing, that is, $\al_k<\dots<\al_{n-1}<\al_0<\dots<\al_{k-1}$. \big(Indeed, in view of Proposition~\ref{prop:preserv}, the reflection transformation $R$ will preserve the convexity property of $\P$ and interchange the property of $\P$ being c-increasing with it being c-decreasing; also, $R$ will preserve the property $\al_0=0$.\big) Then the conditions $\al_0=0$ and $\al_i\in[0,2\pi)$ $\forall i$ imply that $k=0$ and $\al_0=0<\dots<\al_{n-1}$; that is, the sequence $\arg\P$ is increasing.  

Hence, inequality $\al_1\ge\pi$ would imply $\al_i\in(\pi,2\pi)$ for all $i\in\intr2{n-1}$. Hence and because $n\ge3$, one would have
$0=y_1\ge y_2>y_3>\dots>y_n=y_0=0$, and at least one inequality here is strict (since $n\ge3$), which is a contradiction. 

The case $\al_1<\pi$ is similar. In this case, 
$y_2>0$. To obtain a contradiction, suppose that the set $L:=\{i\in\intr2{n-1}\colon y_i\le0\}$ is non-empty and then let $\ell:=\min L$, so that $\ell\in\intr3{n-1}$, $y_{\ell-1}>0$, and $y_\ell\le0$. Then $\al_{\ell-1}\in(\pi,2\pi)$. Hence and because the sequence $\arg\P$ is increasing, one has $\al_i\in(\pi,2\pi)$ for all $i\in\intr{\ell-1}{n-1}$. Therefore,
$0\ge y_\ell>\dots>y_n=y_0=0$, which is a contradiction. 
This contradiction means that $L=\emptyset$, so that $y_i>0$ for all $i\in\intr2{n-1}$; that is, according to \cite[Definition~2.4]{test}, the polygon $\P=(V_0,\dots,V_{n-1})$ is strictly to one side of its edge $[V_0,V_1]$. 

Similarly it is proved that $\P$ is strictly to one side of any other one of its edges; that is, $\P$ is strictly to-one-side. To complete the proof of the ``if'' part of Theorem~\ref{th:angl-test-strict}, it remains to refer to \cite[Lemmas~2.6 and 2.11]{test}.

\textbf{``Only if''}\quad Here is assumed that polygon $\P$ is strictly convex. Again w.l.o.g.\ one has $\al_0=0$. Also, by Remark~\ref{rem:a-text-strict}, w.l.o.g.\ $n\ge4$. Again by the ``reflection'' part of Proposition~\ref{prop:preserv}, w.l.o.g.\ $y_2\ge0$. Moreover, because of the strictness of $\P$ and the assumptions $V_0=(0,0)$ and $V_1=(1,0)$, one actually has $y_2>0$, so that $\al_0=0<\al_1<\pi$ and $\De_{0,1,2}=y_2>0$. 
So, the strict convexity of $\P$ and Remark~\ref{rem:minimality} yield $\De_{0,1,i}=y_i>0$ for all $i\in\intr2{n-1}$. 
The strictness of $\P$ also implies that all the values $\al_0,\dots,\al_{n-1}$ are distinct from one another.
 
Thus, it suffices to show that $\al_i\le\al_{i+1}$ for all $i\in\intr1{n-2}$. Suppose the contrary, that $\al_i>\al_{i+1}$ for some $i\in\intr1{n-2}$. Consider separately the following three cases.

\emph{Case 1: $i=1$.}\quad Then $\al_1>\al_2$. By \eqref{eq:mehl-order-simpl}, this implies that $\De_{1,2,3}\le0$ or $y_2-y_1\le0\le y_3-y_2$; but $y_2-y_1=y_2>0$, so that one must have $\De_{1,2,3}\le0$; now inequalities $\De_{1,2,3}\le0$ and $\De_{0,1,2}>0$ contradict Remark~\ref{rem:minimality}. 

\emph{Case 2: $i=n-2$.}\quad Then $\al_{n-2}>\al_{n-1}$. 
This case is quite similar to Case~1. 
Indeed, by \eqref{eq:mehl-order-simpl}, here one has $\De_{0,n-2,n-1}=\De_{n-2,n-1,0}\le0$ or $0\le y_0-y_{n-1}$; but $y_0-y_{n-1}=-y_{n-1}<0$, so that $\De_{0,n-2,n-1}\le0$; now inequalities $\De_{0,n-2,n-1}\le0$ and $\De_{0,1,2}>0$ contradict Remark~\ref{rem:minimality}. 

\emph{Case 3: $i\in\intr2{n-3}$ \emph{and} $\al_i>\al_{i+1}$.}\quad Then the $5$-gon $\Q:=(V_0,V_1,V_i,V_{i+1},V_{i+2})$ is a sub-polygon of $\P$, so that $\Q$ is strictly convex, by \cite[Corollary 1.17]{elimin}. On the other hand, $\arg\Q=(\al_0,\be,\al_i,\al_{i+1},\ga)$, for some real numbers $\be$ and $\ga$. Thus, w.l.o.g.\ $\P=\Q$, $n=5$, and so, one has all of the following: $\P=(V_0,V_1,V_2,V_3,V_4)$; $i=2$; $\al_2>\al_3$; and $\De_{0,1,i}=y_i>0$ for all $i\in\intr24$. 
By Remark~\ref{rem:minimality}, one now also sees that the determinants $\De_{2,3,4}$, $\De_{0,2,3}$, and $\De_{0,3,4}$ are all strictly positive as well. Therefore, the condition $\al_2>\al_3$ and implication \eqref{eq:mehl-order-simpl} yield $y_3-y_2\le0\le y_4-y_3$. 
One can verify the identity
$$(y_4-y_3)\,\De_{0, 2, 3}+(y_2-y_3)\,\De_{0, 3, 4} + 
\De_{2, 3, 4}\,\De_{0, 1, 3}=0,$$
whose left-hand side 
is strictly positive, since 
all the determinants $\De_{\cdot,\cdot,\cdot}$ in this identity are strictly positive and because of the condition $y_3-y_2\le0\le y_4-y_3$. 
Thus, one obtains a contradiction.  

The proof of the ``only if'' part and thus of entire Theorem~\ref{th:angl-test-strict} is now complete. 
\end{proof}

Let $U\_\_V\_\_W\_\_\dots$ (respectively, $\widehat{U,V,W,\dots}$) mean that $U,V,W,\dots$ are collinear (respectively, non-collinear) points on the plane.

\begin{lemma}\label{lem:calculation}
\emph{[Cf. \cite[Lemma~2.7]{test}.]}\quad
For any choice of $\al$, $\be$, $i$, and $j$ in $\intr0{n-1}$, 
points $V_\al$ and $V_\be$ are to one side of of $[V_i,V_j]$ if and only if
$\De_{\al,i,j}\,\De_{\be,i,j}\ge0$,
where $\De_{\al,i,j}$ are given by \eqref{eq:De}; for an exact definition of ``to one side'', see \cite[Definition~2.4]{test}.
\end{lemma}

\begin{proof}[Proof of Lemma~\ref{lem:calculation}] 
This proof can be done quite similarly to that of \cite[Lemma~2.7]{test}. Alternatively and more simply, Lemma~\ref{lem:calculation} can be easily deduced from \cite[Lemma~2.7]{test} by observing that $V_\al\_\_V_i\_\_V_j$ if and only if $\De_{\al,i,j}=0$.
\end{proof}

\begin{proof}[Proof of Proposition~\ref{prop:strict}]
W.l.o.g., $n\ge4$.

\textbf{(I)$\implies$(II)\quad}
Here it is assumed that 
a polygon $\P=(V_0,\dots,V_{n-1})$ is ordinary and locally-strictly convex. 
By the c-shift invariance (Remark~\ref{rem:c-shift}), 
at this point it suffices to show that $\widehat{V_0,V_1,V_i}$ for each $i\in\intr2{n-1}$ or just only for each $i\in\intr3{n-1}$, because $\widehat{V_0,V_1,V_2}$ follows from $\P$ being locally-strict.


To obtain a contradiction, assume first that $V_0\_\_V_1\_\_V_3$. 
Since $\P$ is locally-strictly convex, one has $\widehat{V_0,V_1,V_2}$, so that, by the affine invariance (Proposition~\ref{prop:preserv}),  
w.l.o.g.\ $V_0=(0,0)$, $V_1=(1,0)$, and $V_2=(1,1)$. Then $V_0\_\_V_1\_\_V_3$ implies that $V_3=(x_3,0)$ for some real $x_3$. Since $\P$ is ordinary, one has $x_3\notin\{0,1\}$. Hence, there are only the following three cases to consider at this point:\\
\emph{Case 1: $x_3>1$}.\quad Then $\De_{1,2,0}\,\De_{1,2,3}=1-x_3<0$. By Lemma~\ref{lem:calculation}, this means that $V_0$ and $V_3$ are not on one side of $[V_1,V_2]$; by \cite[Lemma~2.3]{elimin}, this contradicts the convexity of polygon $\P$. \\
\emph{Case 2: $0<x_3<1$}.\quad Then $\De_{2,3,0}\,\De_{2,3,1}=x_3(x_3-1)<0$, so that $V_0$ and $V_1$ are not on one side of $[V_2,V_3]$, which contradicts the convexity of polygon $\P$. \\
\emph{Case 3: $x_3<0$}.\quad Then 
$V_{n-1}=(x_{n-1},y_{n-1})$ for some real $x_{n-1}$ and $y_{n-1}$ such that $y_{n-1}\ne0$ -- because $\P$ is locally strict and hence $\widehat{V_{n-1},V_0,V_1}$. Hence, $\De_{n-1,0,1}\,\De_{n-1,0,3}\\
=y_{n-1}^2 x_3<0$, so that $V_0$ and $V_3$ are not on one side of edge $[V_{n-1},V_0]$, which contradicts the convexity of polygon $\P$. 

Thus, in all cases the assumption $V_0\_\_V_1\_\_V_3$ leads to a contradiction, so that one has $\widehat{V_0,V_1,V_3}$. 
Similarly one proves that $\widehat{V_1,V_3,V_4}$ (for $n=4$ this has been already proved, for then $V_4=V_0$).
That is, the sub-polygon $\P^{(2)}:=(V_0,V_1,V_3,\dots,V_{n-1})$ is locally strict. Also, 
by Theorem~\ref{cor:sub-polygon},
$\P^{(2)}$ inherits the property of $\P$ of being ordinarily convex. Now it follows by induction that $\P^{(2)}$ is quasi-strict, so that indeed $\widehat{V_0,V_1,V_i}$ for each $i\in\intr3{n-1}$. 
This proves (I)$\implies$(II).

\textbf{(II)$\implies$(III)\quad} This is immediate from \cite[Lemma~2.11]{test}.

\textbf{(III)$\implies$(I)\quad} This is immediate from \cite[Lemmas~2.11 and 2.5]{test}.
\end{proof}

Introduce the ``direction'' equivalence on $\R^2\setminus\{\vec0\}$ defined by the formula
$$\uu\pp\vv\overset{\textrm{def}}\iff
\text{$\vv=\la\uu$ for some $\la>0$},$$
for any $\uu$ and $\vv$ in $\R^2\setminus\{\vec0\}$.
Note that 
\begin{equation}\label{eq:pp}
\uu\pp\vv\implies(\uu+\vv)\pp\uu.	
\end{equation}

\begin{lemma}\label{lem:coll,al}
For any two vectors $\uu$ and $\vv$ in $\R^2\setminus\{\vec{0}\}$, one has $\arg\uu=\arg\vv$ iff $\uu\pp\vv$.
\end{lemma}

\begin{proof}[Proof of Lemma~\ref{lem:coll,al}] 
This folows immediately from \eqref{eq:arg-def}.
\end{proof}

\begin{lemma}\label{lem:collin}
Suppose that a polygon $\P=(V_0,\dots,V_{n-1})$ is locally-ordinary and for some $j$ and $k$ in $\intr0{n-1}$ such that $j\le k$ one has 
$\al_j=\dots=\al_k$, where $(\al_0,\dots,\al_{n-1}):=\arg\P$. Then $\v j{k+1}\ne\vec0$ and 
\renewcommand\theenumi {\emph{(\Roman{enumi})} }
\renewcommand\labelenumi {\theenumi}
\begin{enumerate}
\item
$\arg\v j{k+1}=\arg\v i{i+1}$ for all $i$ in $\intr jk$;
\item
$[V_j,V_{k+1}]=\bigcup_{i=j}^k[V_i,V_{i+1}]$; 
\item 
$[V_i,V_{i+1}]\cap[V_t,V_{t+1}]=\emptyset$ for all $i$ and $t$ in $\intr jk$ such that $i\ne t$.
\end{enumerate}
\end{lemma}

\begin{proof}[Proof of Lemma~\ref{lem:collin}] 
By Lemma~\ref{lem:coll,al}, the condition $\al_j=\dots=\al_k$ implies that
for each $i$ in $\intr0{n-2}$ one has $\v{i+1}{i+2}\pp\v i{i+1}$, so that, in view of \eqref{eq:pp}, $\v j{k+1}\pp\v i{i+1}$, which is equivalent to (I), again by Lemma~\ref{lem:coll,al}.

In particular, these observations imply that the points $V_j,\dots,V_{k+1}$ lie on the same straight line, say $\ell$. Let $A$ be any non-singular affine mapping of $\ell$ onto $\R$ such that $AV_j<AV_{j+1}$ (such a mapping exists because $\P$ is locally-ordinary and hence $V_j\ne V_{j+1}$). 
Then the condition $\v{i+1}{i+2}\pp\v i{i+1}$ implies $V_{i+1}\in(V_i,V_{i+2})$, whence $AV_{i+1}\in(AV_i,AV_{i+2})$, again for each $i$ in $\intr0{n-2}$, so that $AV_j<\dots<AV_{k+1}$, 
$[AV_j,AV_{k+1}]=\bigcup_{i=j}^k[AV_i,AV_{i+1}]$, and 
$[AV_i,AV_{i+1}]\cap[AV_t,AV_{t+1}]=\emptyset$ for all distinct $i$ and $t$ in $\intr jk$.
It remains to note that $[AU,AV]=A[U,V]$ for any points $U$ and $V$ on $\ell$ and recall that $A$ is non-singular and hence one-to-one.
\end{proof}

\begin{lemma}\label{lem:conv then simpl}
Any strictly convex $n$-gon with $n\ge3$ is simple.
\end{lemma}

\begin{proof}[Proof of Lemma~\ref{lem:conv then simpl}] 
Suppose that an $n$-gon $\P=(V_0,\dots,V_{n-1})$ is strictly convex and take any $i$ and $j$ in $\intr0{n-1}$ such that $i<j$. We have then to show that $[V_i,V_{i+1})\cap[V_j,V_{j+1})=\emptyset$. By the c-shift invariance (Remark~\ref{rem:c-shift}), w.l.o.g. $i=0$ and then $j\in\intr1{n-1}$. By Remark~\ref{rem:simpl->ord}, $\P$ is ordinary, so that $V_1\ne V_0$. Let $\ell:=V_0V_1$, the straight line through $V_0$ and $V_1$. By \cite[Lemma~2.6]{test}, there is an open half-plane $H$ such that $\partial H=\ell$ and $\{V_2,\dots,V_{n-1}\}\subset H$. Then (with $i=0$)
\begin{equation}\label{eq:simpl}
[V_i,V_{i+1})\cap[V_j,V_{j+1})
=[V_0,V_1)\cap(\ell\cap[V_j,V_{j+1}))
\subseteq[V_0,V_1)\cap\{V_j\}=\emptyset;	
\end{equation}
the set inclusion in \eqref{eq:simpl} follows because $(V_j,V_{j+1})=\ri[V_j,V_{j+1}]\subseteq\operatorname{interior}(\ell\cup H)=H\subseteq\R^2\setminus\ell$; the last equality in \eqref{eq:simpl} is trivial for $j=1$ and
takes place for $j\in\intr2{n-1}$ because then $[V_0,V_1)\cap\{V_j\}\subseteq\ell\cap H=\emptyset$.
\end{proof}

\begin{lemma}\label{lem:coll then als same}
Suppose that a polygon $\P=(V_0,\dots,V_{n-1})$ is ordinary and locally-simple. Suppose also that for some $j$ and $k$ in $\intr0{n-1}$ one has $V_j\_\_\dots\_\_V_{k+1}$. Then $\al_j=\dots=\al_k$, where $(\al_0,\dots,\al_{n-1}):=\arg\P$.
\end{lemma}

\begin{proof}[Proof of Lemma~\ref{lem:coll then als same}] 
Take any $i$ in $\intr j{k-1}$. Then $\v{i+1}{i+2}=\mu_i\v i{i+1}$ for some $\mu_i\ne0$, given the conditions that  $V_j\_\_\dots\_\_V_{k+1}$ and $\P$ is ordinary. Then inequality $\mu_i<0$ would imply  $[V_i,V_{i+1})\cap[V_{i+1},V_{i+2})\supseteq(\tilde V_i,V_{i+1})\ne\emptyset$, since $\P$ is ordinary, where $\tilde V_i:=V_i$ if $\mu_i\le-1$ and $\tilde V_i:=V_{i+2}$ if $-1<\mu_i<0$; so, this would contradict the condition that $\P$ is locally-simple.

It follows that $\mu_i>0$ and hence $\v{i+1}{i+2}\pp\v i{i+1}$
for all $i$ in $\intr j{k-1}$. It remains to refer to Lemma~\ref{lem:coll,al}.
\end{proof}

\begin{proof}[Proof of Theorem~\ref{th:angl-test-nonstrict}]\ 

\textbf{(I)$\implies$(III)\quad}
Here it is assumed that  
an $n$-gon $\P=(V_0,\dots,V_{n-1})$, with $n\ge3$ and $(\al_0,\dots,\al_{n-1}):=\arg\P$, is ordinary and locally-simply convex. We have to prove that $\P$ is c-monotone.
Introduce the set
$$M:=\{i\in\intr0n\colon\widehat{V_{i-1},V_i,V_{i+1}}\},$$
where 
$V_{-1}:=V_{n-1}$ and $V_{n+1}:=V_1$.

Note that $M\ne\emptyset$. Indeed, otherwise one would have $\al_0=\dots=\al_{n-1}$, by Lemma~\ref{lem:coll then als same}; then Lemma~\ref{lem:collin}(II) would imply $[V_0,V_1]\subseteq[V_0,V_n]=\{V_0\}$, which would contradict the condition that $\P$ is ordinary.

Therefore, $\dim\P=2$. Also, by the c-shift invariance (Remark~\ref{rem:c-shift}), w.l.o.g.\ $0\in M$ and hence (by the definition of $M$) one has $n\in M$,  so that
$$M=\{j_0,\dots,j_m\}$$
for some $m\in\intr0{n-1}$ and integers $j_0,\dots,j_m$ such that $0=j_0<\dots<j_m=n$, and, moreover, 
\begin{equation}\label{eq:^_}
\begin{alignedat}{2}
\widehat{V_{n-1},V_0,V_1},&&\quad V_0=&V_{j_0}\_\_\dots\_\_V_{j_1};\\
\widehat{V_{j_1-1},V_{j_1},V_{j_1+1}}, &&\quad &V_{j_1}\_\_\dots\_\_V_{j_2};\\
& \vdots && \\
\widehat{V_{j_{m-1}-1},V_{j_{m-1}},V_{j_{m-1}+1}}, &\quad  &&V_{j_{m-1}}\_\_\dots\_\_V_{j_m}=V_n=V_0.
\end{alignedat}	
\end{equation}

By Theorem~\ref{cor:sub-polygon}, the sub-polygon
\begin{equation}\label{eq:Q}
	\Q:=(U_0,\dots,U_{m-1}):=(V_{j_0},\dots,V_{j_{m-1}})
\end{equation}
of the ordinarily convex polygon $\P$ is ordinarily convex as well. 

Note also that $\Q$ is locally-strict. To check this, in view of the c-shift invariance (Remark~\ref{rem:c-shift}) it suffices to show that $\widehat{U_0,U_1,U_2}$ or, equivalently, $\widehat{V_{j_0},V_{j_1},V_{j_2}}$. But this follows by the the ordinariness of $\P$ and construction of $\Q$ (whereby $\widehat{V_{j_1-1},V_{j_1},V_{j_1+1}}$, $V_{j_0}\_\_V_{j_1-1}\_\_V_{j_1}$, and $V_{j_1}\_\_V_{j_1+1}\_\_V_{j_2}$).   

Hence, by Proposition~\ref{prop:strict}, $\Q$ is strictly convex. Now, in view of Theorem~\ref{th:angl-test-strict}, $\Q$ is c-strictly monotone. 
Applying (if necessary) the reflection transformation $R$ and a cyclic shift $\th^k$ and referring to Proposition~\ref{prop:preserv} and Remark~\ref{rem:c-shift}, assume w.l.o.g.\ that polygon $\Q$ is increasing:
\begin{equation}\label{eq:Q incr}
	\arg\v{j_0}{j_1}<\dots<\arg\v{j_{m-1}}{j_m}.
\end{equation}
To complete the proof of implication 
(I)$\implies$(III) of Theorem~\ref{th:angl-test-nonstrict}, it remains to refer to \eqref{eq:^_}, Lemma~\ref{lem:coll then als same}, and \eqref{eq:Q incr}, whereby 
\begin{alignat*}{4}
\al_0=& & &&\al_{j_0}&=\dots & &=\al_{j_1-1}=\arg\v{j_0}{j_1} \\
<\arg\v{j_1}{j_2}=& & 
&&\al_{j_1}&=\dots & &=\al_{j_2-1} \\
&&&\vdots&&&& \\
<\arg\v{j_{m-1}}{j_m}=& & 
&&\;\al_{j_{m-1}}&=\dots & &=\al_{j_m-1}=\al_{n-1}.
\end{alignat*}

\bigskip

\textbf{(III)$\implies$(II)}\quad
Here it is assumed that 
a polygon $\P=(V_0,\dots,V_{n-1})$ with $(\al_0,\dots,\al_{n-1}):=\arg\P$ is c-monotone.
We have to prove that $\P$ is simply convex.
Applying the reflection transformation $R$ and a cyclic shift $\th^k$, w.l.o.g.\ polygon $\P$ is non-decreasing:
\begin{equation*}
	\al_0\le\dots\le\al_{n-1}. 
\end{equation*}
That is, 
\begin{equation}\label{eq:al_j}
	\al_{j_0}=\dots=\al_{j_1-1}<\dots<\al_{j_{m-1}}=\dots=\al_{j_m-1} 
\end{equation}
for some natural $m$ and integer $j_0,\dots,j_m$ such that
\begin{equation}\label{eq:js}
	0=j_0<\dots<j_m=n.
\end{equation}
Define polygon $\Q$ again by \eqref{eq:Q}. Then, in view of Lemma~\ref{lem:collin}(I), one has $\arg\Q=(\al_{j_0},\dots,\al_{j_{m-1}})$, so that
$\Q$ is increasing and hence, by Theorem~\ref{th:angl-test-strict}, strictly convex. 

By Lemma~\ref{lem:collin}(II), one has 
$$[U_p,U_{p+1}]=[V_{j_p},V_{j_{p+1}}]=
\bigcup_{i=j_p}^{j_{p+1}-1}[V_i,V_{i+1}]$$
for all $p\in\intr0{m-1}$,
so that
$\edg\Q=\edg\P$. Therefore, moreover,
$$\{V_0,\dots,V_{n-1}\}\subseteq\edg\P=\edg\Q\subseteq\conv\Q\subseteq\conv\P,$$
whence $\conv\P=\conv\Q$.
Thus, the convexity of $\Q$ implies the convexity of $\P$ and $\dim\Q=\dim\P=2$, so that $m\ge3$. 

Let us now show that $\P$ simple. That is, let us take any $i$ and $j$ in $\intr0{n-1}$ such that $i\ne j$ and show that $[V_i,V_{i+1})\cap[V_j,V_{j+1})=\emptyset$. 
By \eqref{eq:js}, there exist $p$ and $q$ in $\intr0{m-1}$ such that $i\in J_p$ and $j\in J_q$, where $J_s:=\intr{j_s}{j_{s+1}-1}$ for all $s\in\intr0{m-1}$.

If $p=q$ then $[V_i,V_{i+1})\cap[V_j,V_{j+1})=\emptyset$ follows from Lemma~\ref{lem:collin}(III) and \eqref{eq:al_j}.

If $p\ne q$ then, by Lemma~\ref{lem:collin}(II),  
$$[V_i,V_{i+1})\cap[V_j,V_{j+1})\subseteq
[U_p,U_{p+1})\cap[U_q,U_{q+1})
=\emptyset,$$ 
the latter equality taking place since $m\ge3$ and the $m$-gon $\Q$ is strictly convex and hence (by Lemma~\ref{lem:conv then simpl}) simple.

This completes the proof of implication 
(III)$\implies$(II) of Theorem~\ref{th:angl-test-nonstrict}.

\bigskip

\textbf{(II)$\implies$(I)}\quad
This implication is the easiest to prove. Indeed, if a polygon $\P$ is simple then it is (trivially) locally-simple and (by Remark~\ref{rem:simpl->ord}) ordinary. 
This completes the proof of Theorem~\ref{th:angl-test-nonstrict}.
\end{proof}

\renewcommand{\refname}{\textsf{\bf Literature}}

\bigskip

{\parskip0pt \parindent0pt \it Department of Mathematical Sciences

Michigan Technological University

Houghton, MI 49931

USA

e-mail: ipinelis@mtu.edu}

\end{document}